\def\lQ{\Lambda_{\rm QCD}}

\newcommand{\be}{\begin{equation}}
\newcommand{\ee}{\end{equation}}
\newcommand{\bea}{\begin{eqnarray}}
\newcommand{\eea}{\end{eqnarray}}

\def\als{\alpha_{\rm s}}
\def\siml{{\ \lower-1.2pt\vbox{\hbox{\rlap{$<$}\lower6pt\vbox{\hbox{$\sim$}}}}\ }}
\def\simg{{\ \lower-1.2pt\vbox{\hbox{\rlap{$>$}\lower6pt\vbox{\hbox{$\sim$}}}}\ }}

\documentclass[allclo]{FBSart}
\usepackage{amsfonts}
\usepackage{amssymb}

\title{Effective Field Theories for Heavy Quarkonium}
\author{N. Brambilla\thanks{\textit{E-mail address:} 
nora.brambilla@ptt.mi.infn.it}}
\institute{
Dipartimento di Fisica dell' Universit\'a di 
Milano and INFN, via Celoria 16,
20133 Milano, Italy}

\runningauthor{N. Brambilla}
\runningtitle{Effective Field Theories for Heavy Quarkonium}
\begin{document}

\maketitle
\begin{abstract}
We briefly review how  nonrelativistic effective field theories  
give us a definition  of  the QCD potentials and a coherent 
field theory derived quantum mechanical scheme  to calculate 
the properties of bound states made  by two or more heavy quarks.
In this framework heavy quarkonium properties depend only on 
the QCD parameters (quark masses and $\als$) and nonpotential 
corrections are systematically accounted for.
 The relation  between the form of the nonperturbative potentials and the low energy QCD 
dynamics is also discussed.
\end{abstract}

\section{The Physical System}
Hadron properties should be obtained from the QCD Lagrangian as a
function  of the coupling constant $\als$ and of the quark masses $m$.
In practice, things are made complicate  by   
QCD being  a strongly coupled theory in the low energy region.
At the scale $\lQ$, nonperturbative effects become dominant and $\als$ becomes large.
The nonperturbative QCD dynamics originates the confinement of quarks inside hadrons.
Typical approaches include, on one hand the use of phenomenological potential models and constituent 
quark model descriptions, on the other hand   first principles lattice simulations (still far from 
the physical parameter window in many cases). 
However, the physics of systems with a heavy quark $Q$ allows some simplification. 
 The quark mass scale $m_Q$ is large,  bigger than $\lQ$.
Then $\als(m_Q)$ is small and perturbative expansions may be performed at this scale.
Bound systems made of two or more heavy quarks
are even more interesting \cite{Brambilla:2004wf}.
 They are nonrelativistic systems characterized by another small parameter, 
the heavy-quark velocity $v$, and by a hierarchy of energy scales: $m_Q$ (hard),
the  relative momentum $p \sim m_Q v$ (soft),
 and the binding energy $E \sim m_Q v^2$ (ultrasoft).
For energy scales close to $\lQ$, perturbation theory breaks down  and one has to rely on nonperturbative 
methods. Regardless of this, the nonrelativistic hierarchy $m_Q \gg m_Q v \gg m_Q v^2$ 
persists also below the $\lQ$ threshold. While the hard scale is always  larger than 
$\lQ$, different  situations may arise for the other two scales 
depending on the considered quarkonium system.
The  soft scale, proportional to the  inverse typical radius $r$,
may be a perturbative ($\gg \lQ$) or a nonperturbative scale ($\sim \lQ$) depending 
on the physical system. The ultrasoft
scale  may still be perturbative only in the case of  $t\bar{t}$ threshold states.

\section{Scales and EFTs: down to pNRQCD}
Taking advantage of  the existence of a hierarchy of scales, 
one can  introduce nonrelativistic effective field theories
(NR EFTs) \cite{Brambilla:2004jw}
 to describe heavy quarkonia.  
A hierarchy of EFTs may be constructed by systematically integrating out 
modes associated to  high energy scales not relevant for quarkonium.
Such integration  is made  in a matching procedure enforcing 
the  equivalence between QCD and the EFT at a given 
order of the expansion in $v$ ($v^2 \sim 0.1$ for $b\bar{b}$,
 $v^2 \sim 0.3$ for $c\bar{c}$,  $v \sim 0.1$ for $t\bar{t}$).
The EFT  realizes a factorization at the Lagrangian level between 
the high energy contributions, encoded into the  matching coefficients, and 
 the low energy contributions, carried by the dynamical degrees of freedom.
Poincar\'e symmetry remains  intact in a nonlinear realization at the level of the NR EFT
and imposes exact relations among the 
matching coefficients \cite{poincare0,Manohar:1997qy}.

By integrating out the hard modes one  obtains Nonrelativistic QCD
 \cite{Caswell:1985ui,Bodwin:1994jh,Manohar:1997qy}.
NRQCD is making explicit at the Lagrangian level the expansions in $m_Q v/m_Q$ and $m_Qv^2/m_Q$.
It is is similar to HQET, but with a different power counting and
accounts also for contact interactions 
between quarks and antiquark pairs (e.g. in decay processes), hence having a wider 
set of operators.
 In NRQCD soft 
and ultrasoft scales are  dynamical and  their mixing  may complicate 
 calculations,  power counting and do not allow to obtain a Schr\"odinger
formulation in terms of potentials. 
One can go down one step further and integrate out  the soft scale 
in a matching procedure to the lowest energy EFT that can be 
introduced for quarkonia,
where only ultrasoft degrees of freedom remain dynamical. Such EFT is called potential 
  NonRelativistic QCD (pNRQCD)   
\cite{Pineda:1997bj,Brambilla:1999xf}.
In this case the matching coefficients encode 
the information on the soft scale and are the potentials.
pNRQCD is making explicit at the Lagrangian level the expansion in $m_Qv^2/mv_Q$.
This EFT  is close to a Schr\"odinger-like description of the bound
state. The bulk of the interaction
is carried by potential-like terms, but non-potential interactions,
associated with the propagation of low-energy degrees of freedom 
($Q\bar{Q}$ colour singlets, $Q\bar{Q}$  colour  octets and low energy gluons),
are generally present. They 
start  to contribute at  NLO 
in the multipole expansion of the gluon fields and are 
typically  related to nonperturbative effects
\cite{Brambilla:1999xf}.

In this EFT frame, 
it is important to establish when $\lQ$ sets in, i.e. when we have to 
resort to non-perturbative methods.
For low-lying resonances, it is reasonable  to assume 
$m_Qv^2 \simg \lQ$. The system is weakly coupled and we may rely on perturbation theory,
for instance, to calculate the potential. In this case, we deal 
with weak coupling pNRQCD.
The theoretical challenge here is 
performing higher-order perturbative calculations 
and the goal is precision physics (see e.g.
\cite{Brambilla:2007pt,Brambilla:2004jw}). 
For higher resonances $m_Qv \sim \lQ$. In this case, we deal with 
strongly coupled pNRQCD. We need nonperturbative methods to calculate the potentials
and one of the goal is the investigation of the QCD low energy dynamics.

\section{The QCD potentials}

pNRQCD \cite{Pineda:1997bj,Brambilla:1999xf}
realizes modern renormalization theory in the context 
of simple nonrelativistic Quantum Mechanics 
\cite{Lepage:1997cs}.
In this framework the Schr\"odinger equation is exactly the equation to be solved 
to get the binding.  The  $Q\bar{Q}$ potentials to be used in such 
equation are the Wilson matching coefficients of pNRQCD 
obtained by integrating out from QCD  all degrees of freedom but the ultrasoft ones.
\leftline{\it Perturbative Potentials}
If the quarkonium system is small, the soft scale is perturbative and the 
potentials can be {\it entirely} calculated in perturbation theory 
\cite{Brambilla:2004jw}. They   undergo renormalization, 
develop a scale dependence and satisfies renormalization
group equations, which eventually allow to resum potentially large logarithms.
Since the degrees of freedom that enter the Schr\"odinger description 
are in this case both $Q\bar{Q}$  color singlet and $Q\bar{Q}$ color octets,
both singlet and octet potentials exist.
At the moment, the  static singlet $Q\bar{Q}$  potential is known at three loops apart from the 
constant term. The first log related to ultrasoft effects arises at three 
loops. Such logarithm  contribution at N$^3$LO 
and the single logarithm contribution at N$^4$LO may be extracted respectively 
from a one-loop and two-loop  calculation in the EFT and have been calculated 
in \cite{Brambilla:1999qa,Brambilla:2006wp}.
The singlet 
static energy,  given by the sum of a constant, the static potential and the ultrasoft 
corrections,
is free from ambiguities of the perturbative series (renormalon). By comparing it
(at the NNLL) with lattice 
calculations of the static potential one sees 
that the QCD perturbative series converges very nicely 
to and agrees with 
the lattice result in the short range and that no nonperturbative
linear (``stringy'') contribution to the static potential exist
\cite{Pineda:2002se,Brambilla:2004jw}. 
\par
The ${1/m_Q}$ singlet potential is known at two loops, the ${1/m_Q^2}$ 
singlet spin-dependent and spin-indepent potentials are known  at one loop
(see the discussion and the original references quoted in  
\cite{Brambilla:2004jw,Kniehl:2002br,Brambilla:1999xj}).

\leftline{\it Nonperturbative potentials}
If the quarkonium system is large, the soft scale is nonperturbative and the 
potentials cannot be  entirely calculated in perturbation theory 
\cite{Brambilla:2004jw}.
They  come out factorized in the product of 
 NRQCD matching coefficients and low energy nonperturbative parts given in terms
of Wilson loops expectation values  and field strengths insertions in the Wilson loop.
The full expression for the QCD potentials up to order $1/m_Q^2$ has been 
obtained in \cite{Brambilla:2000gk}, for the $QQQ$ and $QQq$ case see 
\cite{Brambilla:2005yk}.
Such expressions correct and generalize previous findings in the Wilson loop approach
\cite{Eichten:1980mw} 
that were typically missing the high energy parts of the potentials, 
encoded into the NRQCD matching coefficients and containing the 
dependence on the logarithms of $m_Q$, and some of the low energy contributions.
Poincar\'e invariance establishes exact relations between the potentials
\cite{poincare0} of the type of the  Gromes relation between spin 
dependent and static potentials \cite{Gromes:1984ma}.

In this regime, {\it from pNRQCD we recover the quark potential singlet model}. 
However, here the potentials are calculated from QCD by nonperturbative 
matching. Their evaluation requires calculations on the lattice 
\cite{Bali:1997am} or in QCD vacuum models \cite{Brambilla:1999ja,Brambilla:1998bt}.
Recently lattice calculations have reached a high degree of precision
\cite{Koma:2007jq} and well defined predictions on the  behaviour of the
${1/m_Q}$ and ${1/m_Q^2}$  potentials in the  confining region became  possible.
Such behaviour seems to deviate from a flux tube picture \cite{Brambilla:1998bt}.

Since now precise lattice data on  the long distance behaviour of  the potentials 
up to order ${1/m_Q^2}$ 
are available, one should relate them 
to QCD vacuum model predictions on expectation values of Wilson loops and  
field strengths  Wilson loop insertions. 
These are gauge invariant objects containing information on the field 
distribution between the quarks that go well beyond the area law content.
They  give us an appropriate way to ``measure'' and characterize the confinement 
mechanism.  For this reason  it would be interesting to get predictions on the behaviour of this objects 
also from string theory.

\section{How to obtain the Spectra}

When the soft scale is perturbative the energy levels are given by the 
expectation value of the perturbative potentials, calculated at the 
needed order of the expansion in $\als$, plus nonpotential nonperturbative 
contributions \cite{Voloshin:1978hc,Brambilla:1999xj,Brambilla:2004jw}.
The latter start to contribute to the energy levels at order $m_Q \als^5$.
They are retardation effects and are systematically accounted for in the EFT.
 They  enter  energy levels (and decay widths) in the form of local or nonlocal 
electric and magnetic condensates.  We  still lack a precise 
and systematic knowledge  of such nonperturbative purely glue 
dependent objects. It would be important to have for them 
lattice determinations or data extractions.
Inside pNRQCD it is possible to relate the 
 leading electric and magnetic nonlocal correlators  
to the gluelump masses and to some existing lattice (quenched) determinations 
\cite{Brambilla:2004jw}.  
However, since the nonperturbative contributions  are suppressed in the power 
counting, it is possible to obtain precise determinations of the masses of the
lowest quarkonium resonances with purely perturbative calculations,
in the cases in which the perturbative series is  convergent 
(i.e. after that the appropriate subtractions of renormalons have been
performed) and  large logarithms 
are resummed.  In this framework  power corrections are unambiguously defined.
For a review of the  predictions on the lowest quarkonium resonances obtained in this framework 
see \cite{Brambilla:2004jw,Brambilla:2007pt}.

When the soft scale is nonperturbative the energy levels are given by the
expectation values of the nonperturbative potentials described in the previous 
section. {\it Nonpotential (or retardation) corrections  do not exist in this case}.  
A full phenomenological application, taking into account 
both the NRQCD matching coefficients and the  recent lattice evaluation of the low 
energy part of the potentials,  has not yet been performed and would be needed.

In both cases, the EFT supply us with a proper and well defined quantum  mechanical framework 
to perform systematic calculations of the quarkonium spectrum.
In particular:
\begin{itemize}
\item{} There is a well defined power counting that states  the terms that should be treated 
as (quantum mechanical) perturbation.
\item{} In higher-order calculations, quantum mechanical perturbation theory requires
regularization and renormalization. pNRQCD gives us a well defined and field 
theory derived quantum mechanical framework to calculate perturbative corrections.
In particular, the soft UV divergences in the potential cancel against NRQCD hard
matching coefficients 
\cite{Pineda:1998kn,Brambilla:2004jw}
 and potential UV divergences in quantum mechanical  perturbation theory
cancel against NRQCD  hard matching coefficients 
\cite{Czarnecki:1999mw,Brambilla:2004jw}, 
leaving well behaved and scale independent predictions for physical 
quantities. Then, in this scheme
no divergences arise from e.g. the iteration of the spin-spin potential
in quantum mechanical perturbation theory, as it happens
typically in phenomenological potential model approaches..
\item{} Spectra are function only of the Standard Model parameters.
Conversely one  can use the spectra in order to  extract $\als$ and $m_Q$ \cite{Brambilla:2007pt}.
\end{itemize}

\section{Decays and Transitions}
While the real parts of the pNRQCD matching coefficients give us the potentials,
the imaginary parts give us the inclusive decays widths \cite{Brambilla:2002nu}.
Also transitions may be worked out in pNRQCD. In \cite{Brambilla:2005zw}
  the M1 transition rates for the lowest quarkonia 
resonances has been calculated and a value for $\Gamma (J/\psi \to \gamma \eta_c)$ compatible
with the experimental data has been obtained. 

\skip -0.3truecm

\section{States close to threshold}

The results on the nonperturbative potentials that we have discussed are valid 
away from threshold  and in the case in which hybrids  develop a mass gap of order $\lQ$ 
with respect to singlet  states \cite{Brambilla:2004jw}, as the lattice indicates. 
For states near or above threshold a general systematic EFT approach has still to be developed,
while lattice results on excited resonances are just appearing. 
Most of the existing analyses, especially for  the many new  states discovered recently 
at the B factories, have to rely on a formalism based on potential models and coupled channels,
see e.g.  \cite{Barnes:2007xu} and refs. therein.

\vskip -0.3truecm

\begin{acknowledge} 
Support inside the  European 
Research Training Network FLAVIA{\it net} 
(FP6, Marie Curie Programs, Contract MRTN-CT-2006-035482) is acknowledged.
\end{acknowledge}

\end{document}